\documentclass[preprint,preprintnumbers,amsmath,amssymb]{revtex4}

\keywords{QCD, Jets, Parton Model, Phenomenological Models}
\preprint{LU-TP 03-42}
\preprint{hep-ph/yymmnn}

\usepackage{epsfig}
\usepackage{color}

\usepackage{graphics}
\usepackage{axodraw}
\usepackage{inputenc}
\usepackage{xspace}
\inputencoding{latin1}
\newcommand{\kTpot}[1]{\ensuremath{k_{\perp}^{#1}}}

\def\zslash{\O\xspace}

\begin{document}

\sloppy

\title{Study of the Linked Dipole Chain Model\\
  in heavy quark production at the Tevatron}

\author{A.V.~Lipatov}
  \affiliation{Physical Department, M.V. Lomonosov Moscow State University,
  119992 Moscow, Russia}
  \email{lipatov@theory.sinp.msu.ru}
\author{L.~L\"onnblad}
  \affiliation{Dept.~of Theoretical Physics,
  S\"olvegatan 14A, S-223 62  Lund, Sweden}
  \email{Leif.Lonnblad@thep.lu.se}
\author{N.P.~Zotov}
  \affiliation{D.V.~Skobeltsyn Institute of Nuclear Physics, M.V. Lomonosov
  Moscow State University,
  119992 Moscow, Russia}
  \email{zotov@theory.sinp.msu.ru}

\begin{abstract}
  We present calculations of charm and beauty production at Tevatron
  within the framework of $k_T$-factorization, using the unintegrated
  gluon distributions as obtained from the Linked Dipole Chain model.
  The analysis covers transverse momentum and rapidity distributions
  and the azimuthal correlations between $b$ and $\bar b$ quarks (or
  rather muons from their decay) which are powerful tests for the
  different unintegrated gluon distributions.  We compare the
  theoretical results with recent experimental data taken by D\zslash
  and CDF collaborations at the Tevatron Run I and II.
\end{abstract}

\maketitle

\section{Introduction}
\label{sec:intro}

It is known that in the description of a given cross section in
lepton-proton or proton-proton interactions at high energies it is not
enough to consider only the leading order perturbative terms. Although
at large scales, $\mu$, the running coupling constant $\alpha_s$ may
be small, each power of $\alpha_s$ is accompanied by large logarithms
due to the large phase space available for the additional gluon
radiation.  The solution of this problem is to resum the leading
logarithmic behavior of the cross section to all orders, thus
rearranging the perturbative expansion into a more rapidly converging
series.

The most familiar resummation strategy is based on
Dokshitzer-Gribov-Lipatov-Altarelli-Parisi
(DGLAP)\cite{Gribov:1972ri,Lipatov:1975qm,
Altarelli:1977zs,Dokshitzer:1977sg}
evolution equation, where large logarithmic terms proportional
$\alpha_s^n\ln^n(\mu^2/\Lambda_{\rm QCD}^2)$ are taken into account.
The cross sections can be rewritten in terms of process-dependent hard
matrix elements convoluted with universal parton density functions
which are described by the DGLAP equation. In this way the dominant
contributions come from diagrams where parton emissions in initial
state are strongly ordered in virtuality. This is called collinear
factorization, as the strong ordering means that the virtuality of the
parton entering the hard scattering matrix elements can be neglected
compared to the large scale $\mu$.

DGLAP evolution describes most experimental results from
electron-proton and proton-proton colliders.  By using input parton
densities which are sufficiently singular when $x \to 0$, this
formalism can also account for the strong rise of $F_2$ at small $x$,
as observed at HERA. However, there are problems with the description
of non-inclusive observables such as forward jet production in $ep$
and heavy quark production in $ep$ and $p\bar p$ collisions.

At the energies of the HERA, Tevatron and LHC colliders, the hard
scale $\mu$ of the heavy quark and quarkonium production processes is
large compare to the $\Lambda_{\rm QCD}$ parameter but on the other
hand $\mu$ is much less than the total center-of-mass energy:
$\Lambda_{\rm QCD}\ll \mu \ll \sqrt s$.  Therefore in such case it was
expected that the DGLAP evolution should break down. The situation is
classified as "semihard".

It is believed that at asymptotically large energies (very small
$x\propto\mu^2/s$) the theoretically correct description is given by
the
Balitsky-Fadin-Kuraev-Lipatov(BFKL)\cite{Kuraev:1976ge,Kuraev:1977fs,
  Balitsky:1978ic} evolution equation.  Here large terms proportional
to $\alpha_s^n\ln^n(1/x)$ are taken into account. The BFKL evolution
equation actually predicts a strong power-like rise of $F_2$ at small
$x$. Just as for DGLAP, it is possible to factorize an observable into
a convolution of process-dependent hard matrix elements with universal
parton distributions. But as the virtualities (and transverse momenta)
of the propagating gluons are no longer ordered, the matrix elements
have to be taken off-shell and the convolution made also over
transverse momentum with the unintegrated gluon distribution
$\Phi(x,k_T^2)$. The unintegrated gluon distribution determines the
probability to find a gluon carrying the longitudinal momentum
fraction $x$ and transverse momentum $k_T$.  This generalized
factorization is called
"$k_T$-factorization"\cite{Catani:1991eg,Collins:1991ty}.  If the
terms proportional to $\alpha_s^n\ln^n(\mu^2/\Lambda_{\rm QCD}^2)$ and
$\alpha_s^n\ln^n(\mu^2/\Lambda_{\rm QCD}^2)\ln^n(1/x)$ are also
resummed, then the unintegrated gluon distribution function depends
also on the probing scale $\mu$.

The unintegrated gluon distribution in the proton is a subject of
intensive studies\cite{Anderson:2002cf}. Knowledge of these is in
particular necessary for the description of heavy quark production at
future colliders, such as the LHC.  This quantity depends on more
degrees of freedom than the usual collinear parton density, and is
therefore less constrained by the experimental data. Various
approaches to model the unintegrated gluon distribution have been
proposed.  One such approach, valid for both small and large $x$, has
been developed by Ciafaloni, Catani, Fiorani and Marchesini, and is
known as the CCFM model\cite{Ciafaloni:1988ur,Catani:1990yc,
  Catani:1990sg,Marchesini:1995wr}.  It introduces angular ordering of
emissions to correctly treat gluon coherence effects. In the limit of
asymptotic energies, it is almost equivalent to
BFKL\cite{Forshaw:1998uq,Webber:1998we,Salam:1999ft}, but also similar
to the DGLAP evolution for large $x$ and high $\mu^2$.  The resulting
unintegrated gluon distribution functions depend on two scales, the
additional scale $\bar q$ being a variable related to the maximum
angle allowed in the emission.

The Linked Dipole Chain Model
(LDC)\cite{Andersson:1996ju,Andersson:1998bx} is a reformulation and
generalization of the CCFM model, and agrees with CCFM to leading
double logarithmic accuracy. Also LDC is formulated in terms of
$k_T$-factorization and unintegrated distribution functions, but here
these distributions are essential single-scale dependent quantities.
In this paper we will apply the LDC formalism for the description of
the charm and bottom production processes at Tevatron.

The application of $k_T$-factorization supplemented with the BFKL and
CCFM evolution equations to heavy quark hadroproduction is discussed
in\cite{Levin:1991ry,Ryskin:1996sj,Ryskin:2000bz,Hagler:2000dd,
  Jung:2001rp,Baranov:2000gv,Zotov:2003cb,Baranov:2003cd}.  It was
shown that $b\bar b$ production cross section at Tevatron can be
consistently described using the $k_T$-factorization approach together
with different BFKL or CCFM-like unintegrated gluon distributions. The
NLO pQCD calculations based on the DGLAP evolution scheme
underestimate data by a factor about 2 --
5\cite{Acosta:2002qk,Acosta:2001rz,Abe:1997zt,Abbott:1999wu,Abbott:1999se}.
In a previous paper\cite{Baranov:2003cd} the dependences of the $b$
quark, $B$ meson and their decay muon cross sections on different
forms of the unintegrated gluon distribution was investigated.  It was
found that the properties of different unintegrated gluon distributions
strongly influences the $b\bar b$ or muon-muon azimuthal correlations
since these quantities are sensitive to the relative contributions of
different production mechanisms to the total cross
section\cite{Levin:1991ry,Ryskin:1996sj,Ryskin:2000bz,
  Hagler:2000dd,Jung:2001rp,Baranov:2000gv}.

Based on the above mentioned results here we will use the
$k_T$-factorization approach together with LDC unintegrated gluon
distribution functions\cite{Gustafson:2002jy} for the analysis of the
experimental
data\cite{Acosta:2002qk,Acosta:2001rz,Abe:1997zt,Abbott:1999wu,Abbott:1999se}.
To illuminate the effect of the different contributions we will study
three different versions of LDC unintegrated gluon distributions
presented in\cite{Gustafson:2002jy}. It is interesting to note that
the LDC unintegrated gluon distributions has been fitted to the
inclusive $F_2$ data at HERA, and for heavy quark hadroproduction at
Tevatron we will obtain essentially parameter-free theoretical
results.  We also present our predictions for the differential cross
section of $D^*$, $D^+$ and $D^0$ meson production in $p\bar p$
collisions at Tevatron Run II\cite{Chen:2003qe}.

This article is organized as follows. In Section \ref{sec:ldc} we give
a short review of the CCFM and LDC formalism. In Section \ref{sec:res}
we present the numerical results of our calculations and compare them
with the D\zslash\cite{Abbott:1999wu,Abbott:1999se} and
CDF\cite{Acosta:2002qk,Acosta:2001rz,Abe:1997zt,Chen:2003qe} data.
Finally, in Section \ref{sec:sum}, we give some conclusions.

\section{The CCFM evolution and LDC model}
\label{sec:ldc}

\begin{figure}
\begin{picture}(170,230)(0,0)
\ArrowLine(10,15)(50,15)
\Text(20,25)[]{\large $proton$}
\Line(50,20)(100,20)
\Line(50,15)(100,15)
\Line(50,10)(100,10)
\Line(50,20)(80,40)
\GOval(50,15)(10,7)(0){1}
\Text(60,35)[]{\large $k_{0}$}
\Text(130,40)[]{\large $q_{1}$}
\Line(80,40)(120,40)
\Line(80,40)(95,70)
\Text(80,55)[]{\large $k_{1}$}
\Text(145,70)[]{\large $q_{2}$}
\DashLine(87.5,55)(110,55){2}
\Text(118,55)[]{\large $q'_{1}$}
\Text(90,85)[]{\large $k_{2}$}
\Text(155,100)[]{\large $q_{3}$}
\Line(95,70)(135,70)
\Line(95,70)(105,100)
\DashLine(100,70)(120,80){2}
\Line(105,100)(145,100)
\Line(105,100)(110,130)
\Line(110,130)(150,130)
\Line(110,130)(110,160)
\Line(110,160)(150,160)
\DashLine(120,160)(140,150){2}
\DashLine(130,155)(145,155){2}
\ArrowLine(20,220)(70,200)
\Text(166,160)[]{\large $q_{n+1}$}
\Text(100,145)[]{\large $k_{n}$}
\Text(20,210)[]{\large $lepton$}
\Text(100,185)[]{\large $q_{\gamma}$}
\ArrowLine(70,200)(120,220)
\Photon(70,200)(110,160){3}{5}
\end{picture}
  \caption{\label{FIG1}A fan diagram for a DIS event. The quasi-real
    partons from the initial-state radiation are denoted $q_i$, and
    the virtual propagators $k_i$. The dashed lines denote final-state
    radiation.}  
\end{figure}
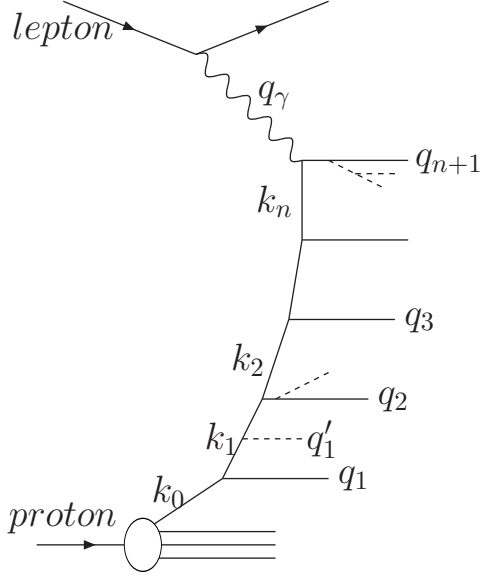

In figure \ref{FIG1} we show a typical fan diagram for the
initial-state radiation in a DIS lepton-proton event. According to the
CCFM formalism the emission of gluons during the initial state cascade
is only allowed in an angular-ordered region of phase space.  In
addition the initial state emissions are ordered in the positive
(along incoming proton) light-cone momentum $k_+$. All other
kinematically allowed emissions (symbolized by $q_1'$ emission in
figure \ref{FIG1}) are defined as final-state emissions.

The CCFM evolution equation can be written in a differential
form\cite{Ciafaloni:1988ur,Catani:1990yc,Catani:1990sg,Marchesini:1995wr}:
\begin{equation}
  \bar q^2{d\over d\bar q^2}
  {\Phi(x,k_T^2,\bar q^2)\over \Delta_s(\bar q^2, Q_0^2)} = 
  \int dz {d\phi\over 2\pi}
    {\tilde P_g(z,(\bar q/z)^2,k_T^2)\over \Delta_s(\bar q^2, Q_0^2)}\,
    \Phi(x',{k_T^2}',(\bar q/z)^2)
  \label{eq:1}
\end{equation}
where $\Phi(x,k_T^2,\bar q^2)$ is the unintegrated gluon distribution,
$\vec k_T = \vec k_T' + \vec q_T$, and $\bar{q}=q_T/(1-z)$ is the
rescaled transverse momentum of an emitted gluon (with azimuthal angle
$\phi$). The Sudakov form factor $\Delta_S$ is given by:
\begin{equation}
  \ln \Delta_s(\bar q^2, Q_0^2) =
  - \int\limits_{Q_0^2 }^{\bar q^2} {dq^2 \over q^2} 
  \int\limits_0^{1 - Q_0/q} dz {\bar \alpha(q^2(1 - z)^2) \over 1 - z}
  \label{eq:2}
\end{equation}
with $\bar \alpha = 3\alpha_S/\pi$. For inclusive quantities at
leading logarithmic order, the Sudakov form factor cancels the $1/(1 -
z)$ collinear singularity of the splitting function. The splitting
function $\tilde P_g$ for branching $i$ is given by:
\begin{equation}
  \tilde P_g(z_i,q_i^2,k_{Ti}^2) = {\bar \alpha(q_i^2(1 - z_i)^2)\over 1 - z_i} +
  {\bar \alpha(k_{Ti}^2)\over z_i} \Delta_{\rm ns}(z_i,q_i^2,k_{Ti}^2)
  \label{eq:3}
\end{equation}
where the non-Sudakov (or non-eikonal) form factor $\Delta_{\rm ns}$
is defined as:
\begin{equation}
  \ln \Delta_{\rm ns}(z_i,q_i^2,k_{Ti}^2) = - \int\limits_{z_i}^1 {dz'\over z'}
  \int {d q^2\over q^2} \bar \alpha \theta (k_{Ti} - q) \theta(q - z'q_i).
  \label{eq:4}
\end{equation}

\noindent The CCFM equation incorporates both BFKL and DGLAP 
evolution. In the small $x$ limit the unintegrated gluon distribution
obeying CCFM evolution equation~(\ref{eq:1}) can be written as:
\begin{equation}
  \displaystyle \Phi(x,k_T^2,\bar q^2) \sim
 \sum_n \int \prod^n \bar \alpha
 {dz_i\over z_i} {d^2 q_{Ti}\over \pi q_{Ti}^2}
 \Delta_{\rm ne}(z_i,k_{Ti}^2,\bar q_i^2)\, \times
 \atop { \displaystyle \times\, \delta(x - \prod z_i)
   \theta (\bar q_i - \bar q_{i-1}z_{i-1}) \delta (k_T^2 - k_{Tn}^2)
   \theta (\bar q - \bar q_n z_n)}
  \label{eq:5}
\end{equation}
where the splitting parameter $z$ is defined as $z_i =
k_{+,i}/k_{+,i-1}$. The interval for $z$ variables is between 0 and 1,
which guarantees ordering in $k_+$, and the angular ordering condition
is satisfied by the constraint
\begin{equation}
  \bar q_i > \bar q_{i - 1}z_{i-1},
  \label{eq:6}
\end{equation}
explicitly written out in eq.~(\ref{eq:5}). The distribution function
depends on two separate scales, the transverse momentum $k_T$ of the
interacting gluon, and $\bar q$, which determines an angle beyond
which there is no (quasi-) real parton in the chain of initial-state
radiation. It may be argued\cite{Jung:2000hk} that the role of this variable
is similar to that of $\mu^2$ in the collinear gluon density.

The LDC model\cite{Andersson:1996ju,Andersson:1998bx} relies on the
observation that the non-Sudakov form factor in equation~(\ref{eq:4})
can be interpreted as a kind of Sudakov giving the no-emission
probability in the region of integration. An additional constraint on
the initial-state radiation is added requiring the transverse momentum
of the emitted gluons to be above the smaller of the transverse
momenta of the connecting propagating gluons:
\begin{equation}
  q_{Ti} > \min (k_{Ti},k_{T i-1}).
  \label{eq:7}
\end{equation}
Emissions failing this cut will instead be
treated as final-state emissions and need to be resummed
in order not to change the total cross section. The
remaining initial-state gluons are ordered both in $q_+$ and
$q_-$, which implies that they are also ordered in angle
or rapidity.

One single chain in the LDC model corresponds to a set of CCFM chains.
It turns out that when one considers the contributions from all chains
of this set, with their corresponding non-eikonal form factors, they
add up to one\cite{Andersson:1996ju}.  Thus, the non-eikonal form
factors do not appear explicitly in LDC, resulting in a simpler form
for the unintegrated gluon distribution function:

\begin{equation}
  \displaystyle \Phi(x,k_T^2) \sim \sum_n \int \prod^n \bar \alpha {dz_i\over z_i} {d^2 q_{Ti}\over \pi q_{Ti}^2} \,\times \atop {
  \displaystyle \times \, \theta(q_{+ i - 1} - q_{+ i}) \theta(q_{- i} - q_{- i - 1}) \delta(x - \prod z_i) \delta (\ln k_T^2 - \ln k_{Tn}^2) }
  \label{eq:8}
\end{equation}

\begin{figure}
  \scalebox{0.9} {\mbox{
\begin{picture}(140,230)(0,-50)
  \Line(10,15)(50,15)
  \Line(50,20)(80,20)
  \Line(50,15)(80,15)
  \Line(50,10)(80,10)
  \Line(50,15)(60,40)
  \GOval(50,15)(10,7)(0){1}
  \Line(60,40)(90,40)
  \Line(60,40)(70,70)
  \Line(70,70)(105,70)
  \Line(70,70)(75,100)
  \Line(75,100)(110,100)
  \Line(75,100)(80,130)
  \Line(80,130)(110,130)
  \Line(80,130)(85,160)
  \Line(85,160)(120,160)
  \Line(85,160)(85,190)
  \Line(85,190)(120,190)
  \Photon(50,210)(85,190){3}{4}
  \Line(85,190)(120,190)
  \Line(20,220)(50,210)
  \Line(50,210)(80,220)
  \Text(47,34)[]{\large $k_{0}$}
  \Text(55,55)[]{\large $k_{1}$}
  \Text(65,85)[]{\large $k_{2}$}
  \Text(70,115)[]{\large $k_{3}$}
  \Text(75,145)[]{\large $k_{4}$}
  \Text(75,173)[]{\large $k_{5}$}
  \Text(100,40)[]{\large $q_{1}$}
  \Text(115,70)[]{\large $q_{2}$}
  \Text(120,100)[]{\large $q_{3}$}
  \Text(125,130)[]{\large $q_{4}$}
  \Text(130,160)[]{\large $q_{5}$}
  \Text(130,190)[]{\large $q_{6}$}
\end{picture}}}\scalebox{0.8} {\mbox{
\begin{picture}(340,280)(0,-20)
  \Line(40,20)(300,20)
  \Line(40,20)(170,280)
  \Line(170,280)(300,20)
  \Line(80,100)(120,20)
  \Text(60,85)[]{$q_{6}$}
  \Vertex(70,80){2}
  \Text(110,70)[]{$k_{5}$}
  \Vertex(125,80){2}
  \Text(125,90)[]{$q_{5}$}
  \Line(90,80)(125,80)
  \Line(125,80)(130,70)
  \Line(130,70)(145,70)
  \Text(140,60)[]{$k_{4}$}
  \Line(145,70)(160,100)
  \Line(160,100)(190,100)
  \Vertex(160,100){2}
  \Text(160,110)[]{$q_{4}$}
  \Text(175,90)[]{$k_{3}$}
  \Vertex(190,100){2}
  \Text(190,110)[]{$q_{3}$}
  \Line(190,100)(200,80)
  \Line(200,80)(230,80)
  \Text(215,70)[]{$k_{2}$}
  \Line(230,80)(250,40)
  \Vertex(230,80){2}
  \Text(230,90)[]{$q_{2}$}
  \Line(250,40)(290,40)
  \Text(270,30)[]{$k_{1}$}
  \Vertex(290,40){2}
  \Text(300,50)[]{$q_{1}$}
  \LongArrow(250,180)(250,230)
  \LongArrow(250,180)(300,180)
  \Text(250,240)[]{$\ln \kTpot{2}$}
  \Text(310,180)[]{$y$}
  \DashLine(125,80)(125,20){2}
  \Line(125,80)(135,10)
  \Line(125,20)(135,10)
  \DashLine(160,100)(160,20){2}
  \Line(160,100)(180,0)
  \Line(180,0)(160,20)
  \DashLine(190,100)(190,20){2}
  \Line(190,100)(210,0)
  \Line(210,0)(190,20)
  \DashLine(230,80)(230,20){2}
  \Line(230,80)(240,10)
  \Line(240,10)(230,20)
  \DashLine(40,-10)(40,20){2}
  \DashLine(120,-10)(120,20){2}
  \DashLine(300,-10)(300,20){2}
  \LongArrow(40,-5)(120,-5)
  \LongArrow(120,-5)(40,-5)
  \LongArrow(120,-5)(300,-5)
  \LongArrow(300,-5)(120,-5)
  \Text(80,-15)[]{$\ln Q^2$}
  \Text(210,-15)[]{$\ln 1/x$}
  \LongArrow(220,130)(260,150)
  \Text(280,150)[]{$\ln q_+$}
  \LongArrow(120,130)(80,150)
  \Text(60,150)[]{$\ln q_-$}
  \DashLine(15,100)(80,100){2}
  \DashLine(15,20)(40,20){2}
  \LongArrow(20,20)(20,100)
  \LongArrow(20,100)(20,20)
  \Text(0,60)[]{$\ln Q^2$}
\end{picture}}}
  \caption{\label{FIG2}The initial-state emissions $q_i$ in the
    $(y,\ln k_T^2)$-plane. Final state radiations is
    allowed in the region below the horizontal lines. The
    height of the horizontal lines determine $\ln k_{Ti}^2$.}
\end{figure}
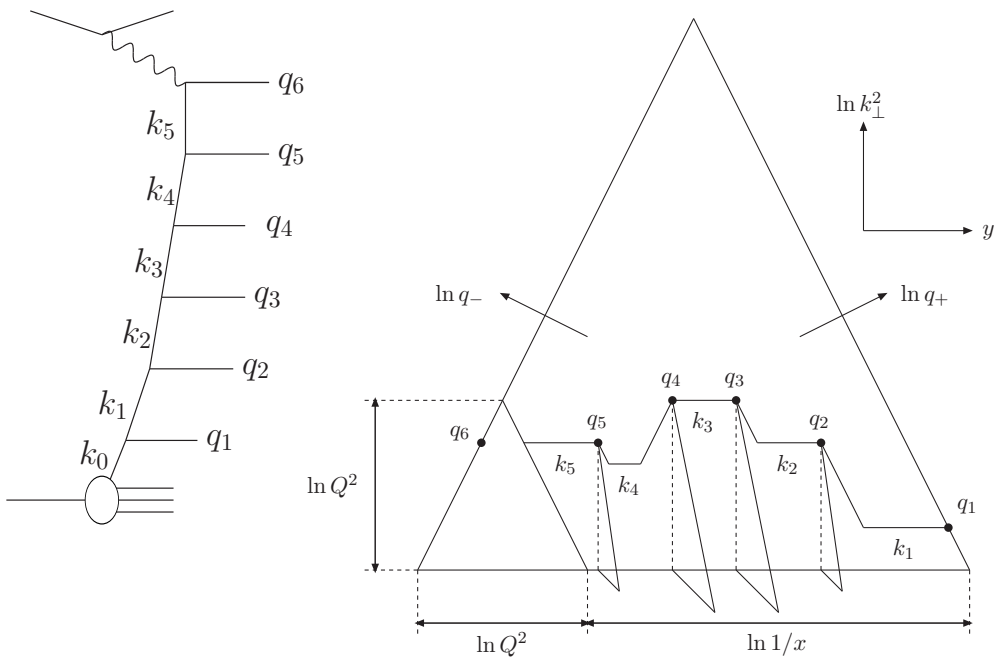

The notation in~(8) and what will follow refers to 
that of figure \ref{FIG2}. Here, a typical DIS event is shown
together with the corresponding phase space available in
the $\gamma^*p$ rest frame, where the rapidity $y$ and 
the transverse momentum $q_T$ of any final-state parton
are limited by a triangular region in the $(y,\ln q_T^2)$-plane.
The proton direction is towards the right end of the triangle,
and the photon direction is towards the left. The real 
emitted gluons are represented by points in this diagram.
The virtual propagators do not have well defined rapidities, and
are represented by horizontal lines, the left and right ends
of which have the coordinates $(\ln k_{+ i}/k_{T i}, \ln k_{Ti}^2)$
and $(\ln - k_{Ti}/k_{-i}, \ln k_{Ti}^2)$ respectively.
The phase space available for final-state emissions is
given by area below the horizontal lines (including the fold
that stick out of the main triangle).

The ordering of the CCFM evolution in $q_+$ but not $q_-$ means that
this formalism is not left-right symmetric. In contrast, the LDC
formulation is completely symmetric, which implies that the chain in
Fig.~2 can be thought of as evolved from either the photon or the
proton end. Thus, the LDC formalism automatically takes into account
resolved virtual photon contributions.

It is important to note that in the LDC model non-leading effects such
as quark-initiated chains, the non-singular terms in the splitting
functions and energy-momentum conservation can be included in a
straight-forward manner. The fact that fewer gluons are considered as
initial-state radiation implies that typical $z$-values are smaller,
thus resulting in smaller sub-leading corrections. It is also implies
that in the LDC formalism the gluon distributions is quite insensitive
to the second scale $\bar q$, and to leading $\ln 1/x$ it depends only
on a single scale $k_T$. Just as for CCFM, for finite $x$ Sudakov form
factors are introduced to regularize the $1/(1-z)$ poles in the
splitting functions, thus introducing a $\bar q$ dependence. This
dependence comes mainly from the last Sudakov form factor related to
the convolution of the unintegrated parton densities with the
off-shell matrix element, and the factorization
\begin{equation}
  \label{eq:9}
  \Phi(x,k_T^2,\bar q^2)=\Delta_s(\bar q^2)\Phi(x,k_T^2)
\end{equation}
is a good approximation for the LDC model.

\section{Numerical results}
\label{sec:res}

In this section we present the numerical results of our calculations
and compare them with data from D\zslash and CDF.  We will use here
the expressions for the heavy quark hadroproduction cross section and
gluon-gluon fusion off-shell matrix element which were obtained
in\cite{Zotov:2003cb,Baranov:2003cd}.

To illuminate the effect of the different contributions we will study
the four different versions of LDC unintegrated gluon distributions
presented in\cite{Gustafson:2002jy}, namely the {\it standard}, {\it
  gluonic}, {\it gluonic-2} and {\it leading} ones.

The {\it standard} version includes non-leading contributions from
quarks and non-singular terms in the splitting functions.  The {\it
  gluonic} and {\it leading} versions do not take into account quark
links in the evolution. Furthermore the {\it leading} version does not
includes non-singular terms in the splitting functions in contrast to
the {\it gluonic} versions.  For all versions, non-perturbative input
parton densities of the form\cite{Kharraziha:1998dn}
\begin{equation}
  xf_i(x,k_{T0}^2) = A_i x^{a_i} (1 - x)^{b_i}
  \label{eq:10}
\end{equation}
where used, with all parameters $A_i$, $a_i$, $b_i$
($i=g,d,\bar{d},u,\ldots$) and the perturbative cutoff $k_{T0}^2$
fitted\cite{Gustafson:2002jy} to reproduce the measured data on
$F_2(x,Q^2)$. The {\it standard} version has been fitted in the region
$x<0.3$, $Q^2>1.5$~GeV$^2$ to experimental data taken by the
H1\cite{Aid:1996au}, ZEUS\cite{Derrick:1996hn},
NMC\cite{Arneodo:1995cq} and E665\cite{Adams:1996gu} collaborations.
The {\it gluonic} and {\it leading} versions have been fitted in the
region $x<0.013$ and $Q^2>3.5$~GeV$^2$ to data taken by the H1
collaboration only.  To study the sensitivity to the $b$ parameter in
equation (\ref{eq:10}) an additional fit for {\it gluonic} case
(called {\it gluonic-2}) was obtained (see\cite{Gustafson:2002jy} for
detail information).  We note that all versions give a satisfactory
fit to the $F_2$ data.

After we fixed the choice of the unintegrated gluon distribution, our
theoretical results depend on the choice of heavy quark mass value,
factorization scale $\mu^2$ and selection of the heavy quark
fragmentation function. In the present paper we convert heavy quarks
into $D$ and $B$ mesons using the usual Peterson fragmentation
function\cite{Peterson:1983ak} with $\epsilon = 0.06$ for charm and
$\epsilon = 0.006$ for bottom respectively\cite{Chrin:1987yd}. Also we
used the following choice $m_c = 1.5\,{\rm GeV}$, $m_b = 4.75\,{\rm
  GeV}$ and $\mu^2 = m_{T}^2 = m_Q^2 + p_{T}^2$ with $p_{T}$ being the
transverse momentum of the heavy quarks in the $p\bar p$ c.m. frame.
These choices are similar to the ones in e.g.\ \cite{Cacciari:2002pa}.

\begin{figure}
  \epsfig{figure=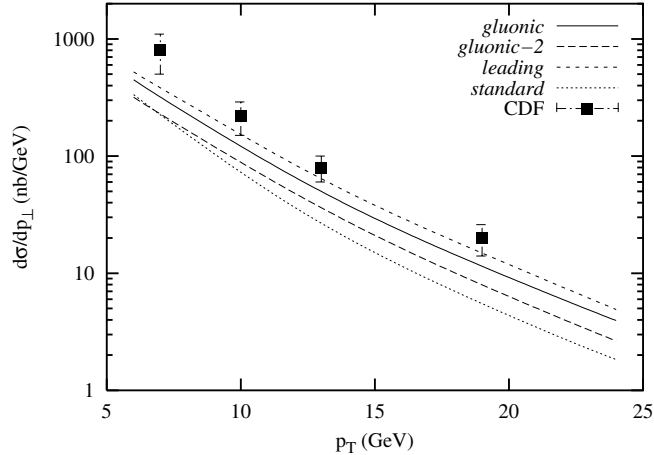,width=9cm}
  \caption{\label{FIG3}The prediction for the $B$ meson $p_T$ spectrum
    for $|y_B|<1$ at $\sqrt{s}=1800$~GeV compared to the CDF data.
    The solid line is {\it gluonic}, dashed is {\it glionic-2}, short
    dashed is {\it leading} and dotted line is {\it standard}
    versions.  Experimental data are from CDF\cite{Acosta:2002qk}.}
\end{figure}

\begin{figure}
  \epsfig{figure=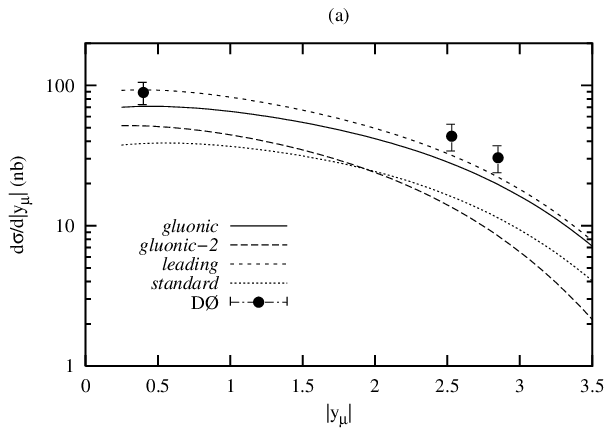,width=8cm}
  \epsfig{figure=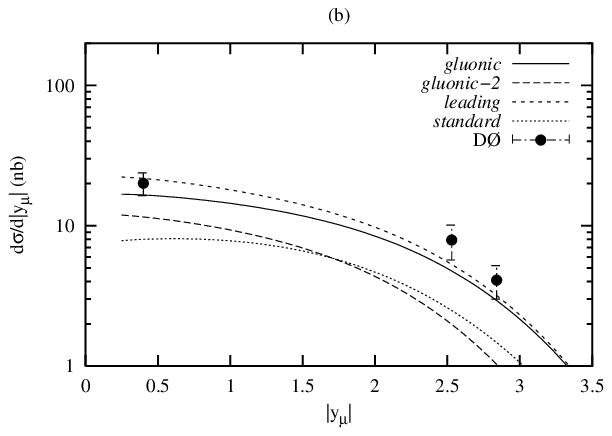,width=8cm}
  \caption{\label{FIG4}The cross section for muons with $p_{T\mu} > 5$~GeV (a)
    and $p_{T\mu} > 8$~GeV (b) from $B$ meson decay as a function of
    rapidity compared to the D\zslash data.  All curves are the same
    as figure \ref{FIG3}.  Experimental data are from
    D\zslash\cite{Abbott:1999se}}
\end{figure}

\begin{figure}
  \epsfig{figure= 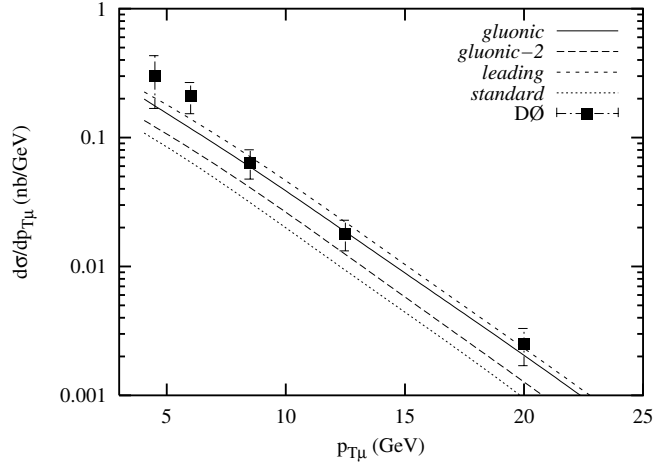,width=9cm}
  \caption{\label{FIG6}The leading muon $p_T$ spectrum for $b\bar b$ production
    compared to the D\zslash data. The cuts applied to both muons are
    $4<p_{T\mu}<25$~GeV, $|\eta_{\mu}|<0.8$ and
    $6<m_{\mu\mu}<35$~GeV. All curves are the same as in figure
    \ref{FIG3}.  Experimental data are from
    D\zslash\cite{Abbott:1999wu}.}
\end{figure}

\begin{figure}
\epsfig{figure=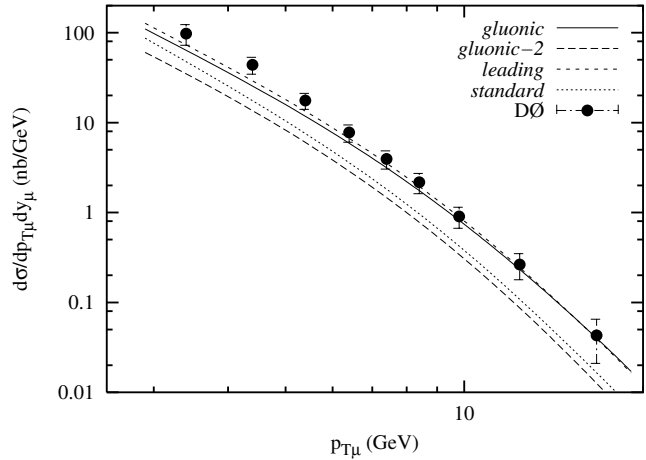,width=9cm}
  \caption{\label{FIG7}The double differential cross sections for muons from
    $B$ meson decay with $2.4 < |y_{\mu}| < 3.2$ as a function of
    $p_{T\mu}$. All curves are the same as in figure \ref{FIG3}.
    Experimental data are from D\zslash\cite{Abbott:1999se}.}
\end{figure}

\begin{figure}
  \epsfig{figure=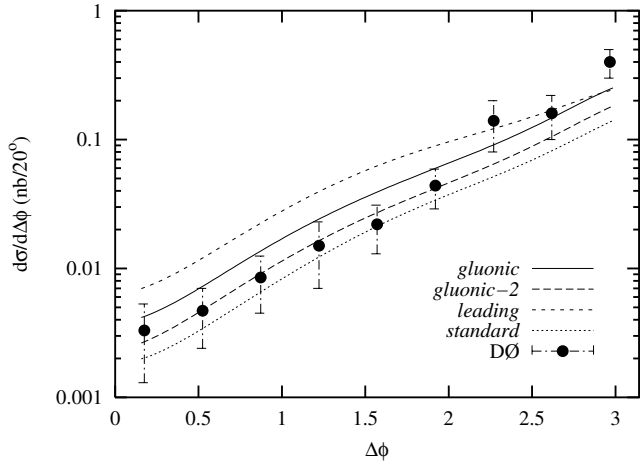,width=9cm}
  \caption{\label{FIG8}Azimuthal correlations for muons pairs with
    $4<p_{T\mu}<25$~GeV, $|\eta_{\mu}|<0.8$ and $6<m_{\mu\mu}<35$~GeV.
    All curves are the same as in in figure \ref{FIG3}.  Experimental
    data are from D\zslash\cite{Abbott:1999wu}.}
\end{figure}

The results of our calculations for bottom production are shown in
figures \ref{FIG3}--\ref{FIG8}.  Figure \ref{FIG3} shows the
prediction for the $B^+$ meson $p_T$ spectrum for $|y_{B}|<1$ at
$\sqrt{s}=1800$~GeV compared to the CDF data\cite{Acosta:2001rz}. One
can see that results obtained with the {\it leading} version agree
with the CDF data within experimental uncertainties. The {\it gluonic}
version is very close to the {\it leading} one but goes a bit below
the data.  The results obtained using the {\it gluonic-2} and {\it
  standard} versions underestimate experimental data by a factor about
2 and are close to the NLO pQCD calculations\cite{Acosta:2001rz}.  One
can also see a difference in the shapes between the {\it standard}
version and the other ones, while the shapes of the {\it gluonic} and
{\it leading} curves practically coincide.

The recent D\zslash\cite{Abbott:1999wu,Abbott:1999se} experimental
data refer also to muons which originate from the semileptonic decays
of $B$-mesons. To produce muons from $B$ mesons in theoretical
calculations, we simulate their semileptonic decay according to the
standard electroweak theory.  Our calculation of the cross section for
muons from $B$ meson decay as a function of rapidity
$d\sigma/d|y_{\mu}|$ is shown in figure \ref{FIG4} for both
$p_{T\mu}>5$~GeV (\ref{FIG4}a) and $p_{T\mu}>8$~GeV (\ref{FIG4}b). We
find that only {\it leading} and {\it gluonic} versions agree with the
D\zslash experimental data\cite{Abbott:1999se} within errors. Again,
the {\it leading} is somewhat above the {\it gluonic} although we now
see that the difference due to the non-singular terms is mainly
present in the central rapidity region.  Also it is interesting to
note that {\it standard} version have a more flat behavior compared to
the other versions of the LDC unintegrated gluon distributions. The
reason that the {\it gluonic-2} results falls much faster at large
rapidities is that it has a more rapidly falling input distribution at
large $x$ ($b_g=7$ in equation (\ref{eq:10}) rather than $b_g=4$ as
for the other versions). Note that the NLO pQCD calculations
underestimate data by a factor about 4\cite{Abbott:1999se}.

Figure \ref{FIG6} shows the leading muon $p_T$ spectrum in the
central rapidity region for $b\bar b$ production compared to the
D\zslash data\cite{Abbott:1999wu}. The leading muon in the event is
defined as the muon with highest transverse momentum. Again we find
that the {\it leading} and {\it gluonic} versions agree well with the data
while the other two are below.  However, the shape of all curves
practically coincide.

Also the double differential cross sections
$d\sigma/dp_{T\mu}\,dy_{\mu}$ in the forward rapidity region
$2.4<|y_{\mu}|<3.2$ are well described by the $k_T$-factorization
approach with the {\it leading} and {\it gluonic} unintegrated gluon
distributions (figure \ref{FIG7}).  Again there is only slight
variations in the shape of the different cureves. The NLO pQCD
calculations underestimate the D\zslash data\cite{Abbott:1999se} by a
factor of 4.

In a previous paper\cite{Baranov:2003cd} it was found that
investigations of $b\bar{b}$ correlations such as the azimuthal
opening angle between $b$ and $\bar{b}$ quarks (or between their decay
muons) is a powerful test for the different unintegrated gluon
distributions. This is because these quantities are sensitive to the
relative contributions of different production mechanisms to the total
cross section\cite{Levin:1991ry,Ryskin:1996sj,Ryskin:2000bz,
  Hagler:2000dd,Baranov:2000gv}.  In the naive collinear gluon-gluon
fusion mechanism, the distribution over the azimuthal angle difference
$\Delta \phi_{b\bar{b}}$ must be a simple delta function
$\delta(\Delta\phi_{b\bar{b}}-\pi)$.  Taking into account
non-vanishing initial gluon transverse momenta leads to the violation
of this back-to-back quark production kinematics in the
$k_T$-factorization approach.

The differential $b\bar{b}$ cross section
$d\sigma/d\Delta\phi_{\mu\mu}$ for central muons is shown in figure
\ref{FIG8}. For the overall cross section we se the same trend as in
previous figures, where the \textit{leading} is above the
\textit{gluonic} which in turn is above the \textit{standard}.
However, we here see from the shape that the \textit{leading} clearly
overestimates the decorrelation. The others curves agree better with
the shape of the experimental result, although none of them are able
to fully reproduce the peak at $\Delta\phi_{\mu\mu}\sim\pi$.

Very recently the CDF collaboration have reported preliminary
experimental data\cite{Chen:2003qe} on charm production at the
Tevatron Run II.  These data found to be about a factor of 1.5 larger
than NLO pQCD theoretical predictions\cite{Cacciari:2003zu}.
data\cite{Chen:2003qe}.

\begin{figure}
  \epsfig{figure=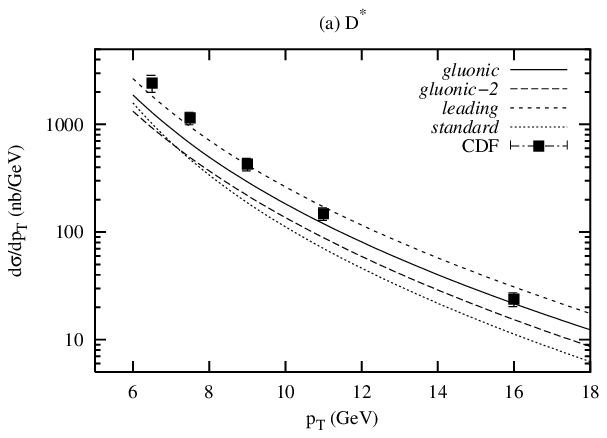,width=8cm}
  \epsfig{figure=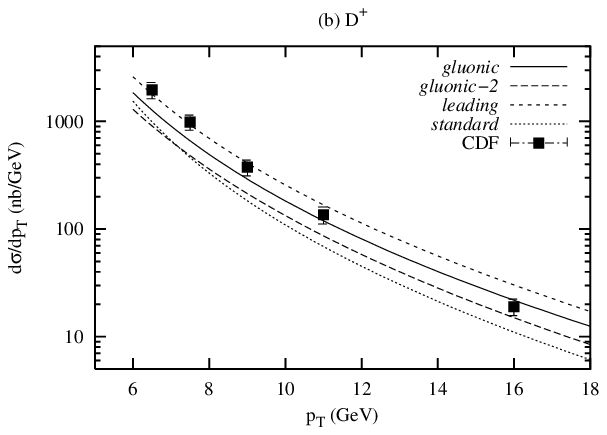,width=8cm}
  \epsfig{figure=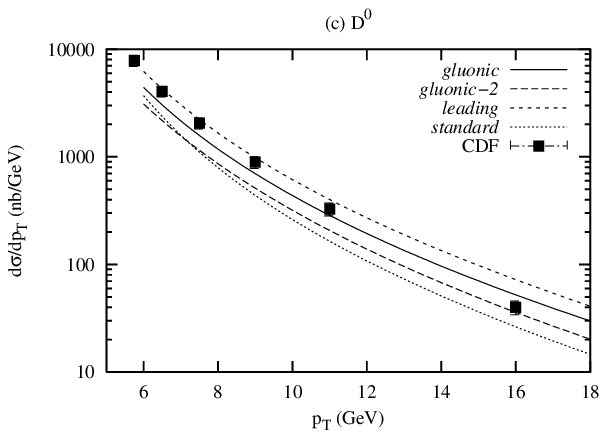,width=8cm}
  \caption{\label{FIG9}The prediction for the $p_T$ spectrum of $D^*$ (a),
    $D^+$ (b) and $D^0$ (c) mesons with $|y_D|<1$ at
    $\sqrt{s}=1960$~GeV compared to the CDF data.  All curves are the
    same as in figure \ref{FIG3}.  Experimental data are from
    CDF\cite{Chen:2003qe}.}
\end{figure}

The results of our calculations of the transverse momentum spectra for
centrally produced $D^*$, $D^+$ and $D^0$ mesons at
$\sqrt{s}=1960$~GeV are shown in figure \ref{FIG9}. Comparing with the
B meson spectrum in figure \ref{FIG3}, we see that the difference
between the different versions of the LDC densities are the same both
for the magnitude (i.e.\ \textit{leading} $>$ \textit{gluonic} $>$
\textit{gluonic-2} $>$ \textit{standard}) and for the shape
(\textit{standard} falls more steeply than the others). However, the
data is now a bit closer to the calculations, especially for the
\textit{leading} version.

\section{Conclusions}
\label{sec:sum}

We have presented results for charm and bottom production in
$p\bar{p}$ collisions at high energies within the framework of
$k_T$-factorization, using different unintegrated gluon densities
obtained from the LDC model. It has previously been noted that the
standard collinear factorization approach fails to describe the amount
of heavy quarks produced at the Tevatron, and that $k_T$-factorization
may be the more correct way of describing the underlying physics. Our
results agrees with this observation. However, within the LDC model it
is also possible to study non-leading effects in the evolution, and it
is clear that these introduce large uncertainties. We have found that
the inclusion of quarks in the evolution has a big effect and that the
results are sensitive to the treatment of non-singular terms in the
gluon splitting function. Also we have found that the results are
sensitive to the assumed shape of the non-perturbative input density at
large $x$.

In particular we find that we can only get an acceptable result for
the case where only gluons are considered in the evolution and where
only the singular terms in the gluon splitting function are included
(\textit{leading}).  This is consistent with the findings in
\cite{Jung:2001rp} where the CASCADE Monte Carlo\cite{Jung:2000hk} was
shown to reproduce the amount of bottom production at the Tevatron.
This program is also based on the CCFM equation and implements purely
gluonic evolution without considering non-leading contributions in the
splitting function. It is interesting to note that for the forward jet
rates at HERA the situation is similar, as these can be reproduced by
LDC and CASCADE only if non-singular terms are
omitted\cite{Anderson:2002cf}.

Although the \textit{leading} version of the LDC unintegrated gluon
density gives a good description of the overall rate of heavy quark
production, less inclusive observables such as azimuthal correlations
between $b\bar{b}$ (or $\mu^+\mu^-$) pairs, are not as well described.

Besides the uncertainties due to non-leading terms in the evolution,
there is, of course, also some uncertainties due to the chosen values
of the heavy quark masses. However, the effects of the heavy quark
masses would not be large enough to change the conclusions presented
here. In addition, there is some freedom in the choice of
factorization scale. We have here chosen $\mu=m_T$ which is the
natural scale within the LDC model, but it can be noted that in e.g.\ 
\cite{Baranov:2000gv}, the scale was chosen to be $\mu=q_T$.  Using
such a scale in the LDC model would increase the bottom cross section,
but, again, it would not be large enough to affect our conclusions.

It is important to note that we have here only considered gluon-gluon
fusion into heavy quarks. One may also expect non-negligible
contribution from two other production mechanisms, namely gluon
splitting, where a hard final-state gluon splits into a heavy quark
pair in a subsequent final-state cascade, and heavy quark excitation
processes, where the the hard process is $gb\rightarrow gb$, with the
$b$ produced by initial-state gluon splitting earlier on in the
evolution. In principle, all these contributions can be taken into
account in the LDCMC Monte Carlo\cite{Kharraziha:1998dn} which
implements the LDC model, however, this program is not yet able to
fully handle hadron-hadron collisions\cite{Gustafson:2002kz}.

To conclude, it is most likely that $k_T$-factorization holds the key
to understanding heavy quark production at the Tevatron. However,
there are still large uncertainties, and much more work needs to be
done before these processes are fully understood.

\section{Acknowledgments}

The study was supported in part by RFBR grant N$^{\circ}$
02--02--17513 and the Crawfoord Foundation (Sweden). A.L.\ was
supported by INTAS grant YSF 2002 N$^{\circ}$ 399.

\bibliography{references}

\end{document}